\documentclass[amsmath,amssymb,aps,showpacs,twocolumn,prb,floatfix]{revtex4}
\usepackage{graphicx}
\usepackage{dcolumn}
\usepackage{bm}
\usepackage{color}
\usepackage{multirow}
\usepackage{times}

\newcommand{\vc}[1]{\mathbf{#1}}
\newcommand{\sqa}{\ensuremath{\hat{s}}}
\newcommand{\EF}{\ensuremath{E_{\rm F}}}
\newcommand{\bra}[1]{\ensuremath{|#1\rangle}}
\newcommand{\up}{\ensuremath{\uparrow}}
\newcommand{\dn}{\ensuremath{\downarrow}}

\begin{document}

\title{Spin relaxation and the Elliott-Yafet parameter in W(001)
  ultrathin films: surface states, anisotropy and oscillation effects}

\author{Nguyen H. Long} 
\email{h.nguyen@fz-juelich.de}
\affiliation{Peter Gr\"unberg Institut and Institute for Advanced Simulation, 
Forschungszentrum J\"ulich and JARA, D-52425 J\"ulich, Germany}
\author{Phivos Mavropoulos}\email{ph.mavropoulos@fz-juelich.de} 
\author{ Bernd Zimmermann, Swantje Heers, David S. G. Bauer, Stefan Bl\"ugel} 
\affiliation{Peter Gr\"unberg Institut and Institute for Advanced Simulation, 
Forschungszentrum J\"ulich and JARA, D-52425 J\"ulich, Germany}
\author{ Yuriy Mokrousov} 
\affiliation{Peter Gr\"unberg Institut and Institute for Advanced Simulation, 
Forschungszentrum J\"ulich and JARA, D-52425 J\"ulich, Germany}
\date{\today}

\begin{abstract}
  Using first-principles methods based on density-functional theory we investigate the spin relaxation in W(001) ultrathin films. 
Within the framework of the Elliott-Yafet theory we calculate the spin mixing of the Bloch states and we explicitly consider spin-flip scattering off self-adatoms. 
We find an oscillatory behavior of the spin-mixing parameter and relaxation
  rate as a function of the film thickness, which we trace back to
  surface-state properties. We also analyze the Rashba effect
  experienced by the surface states and discuss its influence on the
  spin relaxation. Finally we calculate the anisotropy of the 
  spin-relaxation rate with respect to the polarization direction of the
  excited spin population relative to the crystallographic axes of the
  film. We find that the spin-relaxation rate can increase by as much as
  47\% when the spin polarization is directed out of plane, compared
  to the case when it is in plane. Our calculations are based on the
  multiple-scattering formalism of the Korringa-Kohn-Rostoker
  Green-function method.
\end{abstract}

\pacs{72.25.Rb, 73.50.Bk, 72.25.Ba, 85.75.-d}
%\pacs{Valid PACS appear here}

\maketitle

\section{Introduction \label{sec:introduction}}

In the process of spin relaxation an excited electron spin population
returns to the state of equilibrium that, in non-magnetic materials,
corresponds to zero spin polarization.  Despite the fact that the
fundamental mechanisms contributing to spin relaxation have been
investigated since a long time in various systems, the phenomenon
still attracts attention, owing to its importance in spintronics
applications,\cite{zutic04} for example in giant magnetoresistance or,
lately, in the spin Hall effect
as well as the inverse spin Hall effect that is used to probe spin currents.\cite{SHE} 
We mention these examples among a variety of applications in order to stress the practical importance of spin relaxation in thin metallic films, which is part of the motivation for the present work, as a source of loss of spin-mediated information.

There are various mechanisms that can contribute to spin
relaxation\cite{elliott54,yafet63,DyakonovPerel,fabian07,zhukov08,fabian10}
and in metallic systems they are mostly related to Fermi-surface
properties. 
Although it is clear that in a non-magnetic metal or
metallic film the spin relaxation can be attributed to spin-flip
scattering in the presence of the spin-orbit coupling (SOC), many
parameters come into play, and one expects the spin-relaxation rate to
depend strongly on the film's crystallographic orientation, its
thickness and details of the Fermi surface. 
Of particular importance here\cite{heers11} can also be the Rashba states created as a result of spin-orbit interaction acting on the surface bands in the
film.\cite{lashell96,henk03}

In this work we present a study of free-standing W(001) films with
space-inversion symmetry where we consider that the spin-flip scattering is induced by W adatom impurities. 
The motivation for restricting the investigation to free-standing films is
that we consider them to be generic prototypes for films in layered
structures if the material ``sandwiching'' the film --- the
surrounding matrix --- is insulating. There is no question that
contact at the film surfaces will produce effects that depend on the
surrounding matrix, however, we are searching here for physical
mechanisms that can in principle still be present in the contact case,
even if the details of the band structure are different. In the present
analysis we will face e.g.\ oscillatory effects arising from the
thickness dependent interaction of surface states that can be replaced
by interface states if the films are sandwiched; also, the low
dimensionality entails an anisotropy of spin relaxation that should be
a quite general effect irrespective of the film contact. Due to the pronounced manifestation of such effects, bcc W(001) films are chosen among $5d$ transition-metal thin films for a deeper analysis in the present work. 

In the presence of structural inversion symmetry, which is the case here in bcc(001) films, the Elliott--Yafet
mechanism\cite{elliott54,yafet63} plays the most important role for
spin relaxation. Within this mechanism, the relaxation is realized via
spin-orbit mediated spin-flip scattering off impurities at low temperatures and
additionally off phonons at higher temperatures. According to
Elliott,\cite{elliott54} in a system with time-reversal\cite{kramers30} and
space-inversion symmetry there are two degenerate Bloch states at each
{\bf k}-point, which can be written as superpositions of up
$|\uparrow\rangle$ and down $|\downarrow\rangle$
spinors:
\begin{equation}
\begin{array}{lll}
\Psi^{+}_{\vc{k}}(\vc{r})&=&
\left(a_{\vc{k}}(\vc{r})|\uparrow\rangle+b_{\vc{k}}(\vc{r})|\downarrow\rangle\right)
e^{i\vc{k}\cdot\vc{r}} \\
\Psi^{-}_{\vc{k}}(\vc{r})&=&
\left(a^*_{\vc{-k}}(\vc{r})|\downarrow\rangle-b^*_{\vc{-k}}(\vc{r})
|\uparrow\rangle\right)e^{i\vc{k}\cdot\vc{r}}.
\end{array}
\label{mixwavefunction}
\end{equation}
Due to the degeneracy there is an arbitrariness in the selection of
$a_{\vc{k}}(\vc{r})$ and $b_{\vc{k}}(\vc{r})$, as any superposition of the states in
Eq.~(\ref{mixwavefunction}) is also an eigenstate of the Hamiltonian. In practice the
arbitrariness is usually\cite{heers11,Pientka12} lifted by the demand
that the spin expectation value of $\Psi^{+}_{\vc{k}}$ should be
maximal in the $z$ direction, or actually in any chosen
direction\cite{Zimmermann12,mokrousov12} $\sqa$ (that we call the spin-quantization axis, SQA)
experimentally defined by the polarization direction of the excited spin 
population. The spin polarization vector $\vc{S}^{\pm}_{\vc{k}}$
corresponding to $\Psi^{\pm}_{\vc{k}}$ can be calculated via
\begin{equation}
\vc{S}^{\pm}_{\vc{k}}=\frac{1}{2}\langle\Psi^{\pm}_{\vc{k}}\left|\boldsymbol{\sigma}\right|
\Psi^{\pm}_{\vc{k}}\rangle,\ \ \ \mbox{$i=x, y, z$},
\label{spinexpectation}
\end{equation}
where $\boldsymbol{\sigma}$ is the vector of Pauli matrices (atomic
units with $\hbar=1$ are implied).
The two degenerate wavefunctions at each $\vc{k}$-point are orthogonal to each other and have opposite spin expectation values:
\begin{equation}
\vc{S}^{-}_{\vc{k}} = -\vc{S}^{+}_{\vc{k}}
\label{linearcomb}
\end{equation}
It is straightforward to show the relation of the spin projection along the SQA,
$\sqa\cdot\vc{S}^{+}_{\vc{k}}$, to $a_{\vc{k}}(\vc{r})$ and
$b_{\vc{k}}(\vc{r})$:
\begin{eqnarray}
a_{\vc{k}}^2 := \int \left|a_{\vc{k}}(\vc{r})\right|^2\,d^3r&=&\frac{1}{2}+\sqa\cdot\vc{S}^{+}_{\vc{k}},\\
b_{\vc{k}}^2 :=\int \left|b_{\vc{k}}(\vc{r})\right|^2\,d^3r&=&\frac{1}{2}-\sqa\cdot\vc{S}^{+}_{\vc{k}},
\label{elliottyafetpara}
\end{eqnarray}
where we have also introduced the integrals $a_{\vc{k}}^2$ and
$b_{\vc{k}}^2$.
Taking an average over the Fermi surface (FS) (actually Fermi lines in the present two-dimensional case), the Elliott-Yafet parameter (EYP) is obtained as
\begin{equation}
b^2=\left<b_{\vc{k}}^2\right>_{\rm FS}=\frac{1}{n(\EF)V_{\rm BZ}}
\int_{\rm FS}dk\frac{b_{\vc{k}}^2}{v_{\rm F}(\vc{k})},
\label{bsq}
\end{equation}
where $v_{\rm F}({\vc{k}})$ is the Fermi velocity, $n(\EF)$ is the
density of states at the Fermi level $\EF$ and $V_{\rm BZ}$ is the
Brillouin zone volume.  

Momentum scattering events couple the spin-up and down components of the
wavefunctions at different momenta, allowing for transitions between
$\Psi^+_{\vc{k}}$ and $\Psi^-_{\vc{k}'}$ and giving rise to spin
relaxation. In the Elliott approximation\cite{elliott54}
the spin-flip probability rate of state $\Psi^+_{\vc{k}}$,
$P^{+-}_{\vc{k}}$, is proportional to the momentum-dependent
spin-mixing parameter $b_{\vc{k}}^2$.  As a result, after taking an average over the Fermi surface, the ratio
between the spin-relaxation rate $1/T_1$ and momentum-relaxation rate
$1/T_{\rm p}$, $T_1^{-1}/T_{\rm p}^{-1}$, is proportional to $b^2$.
Departing from the Elliott approximation, in which  $b^2$ is assumed small,
one has to take into account the form of the scattering potential $\Delta V$
of the impurity (W adatom in this case) and calculate the scattering $T$-matrix, $T(E)=\Delta V (1 - G(E) \Delta
V)^{-1}$, where $G(E)$ is the Green function of the unperturbed
film.  
Then, the spin-conserving and spin-flip probability rates are
given by the squared matrix elements
$P^{++}_{\vc{k}\vc{k}'}=2\pi|\langle
\Psi^{+}_{\vc{k}}|T(\EF)|\Psi^{+}_{\vc{k}'}\rangle|^2$ and
$P^{+-}_{\vc{k}\vc{k}'}=2\pi|\langle
\Psi^{+}_{\vc{k}}|T(\EF)|\Psi^{-}_{\vc{k}'}\rangle|^2$,
respectively. 
Fermi-surface integrals of the scattering probability
yield the average momentum- and spin-relaxation rate per
$\vc{k}$-point
\begin{eqnarray}
T_1^{-1}(\vc{k})=2T_{\rm sf}^{-1}(\vc{k})&=&\frac{c}{V_{\rm BZ}}\int_{\rm FS} dk' \frac{P^{+-}_{\vc{k}\vc{k}'}+P^{-+}_{\vc{k}\vc{k}'}}{v_{\rm F}(\vc{k}')} \label{eq:T1k}\\
T_{\rm p}^{-1}(\vc{k})&=&\frac{c}{V_{\rm BZ}}\int_{\rm FS} dk'
\frac{P^{++}_{\vc{k}\vc{k}'}+P^{+-}_{\vc{k}\vc{k}'}}{v_{\rm
    F}(\vc{k}')} 
\label{eq:tpk}
\end{eqnarray}
and $\vc{k}$-averaged
\begin{eqnarray}
T_1^{-1}=2T_{\rm sf}^{-1}&=&\frac{1}{n(\EF)V_{\rm BZ}}\int_{\rm FS} dk \frac{T_1^{-1}(\vc{k})}{ v_{\rm F}(\vc{k})}\\
T_{\rm p}^{-1}&=&\frac{1}{n(\EF)V_{\rm BZ}}\int_{\rm FS} dk \frac{T_{\rm p}^{-1}(\vc{k})}{ v_{\rm F}(\vc{k})}
\label{eq:tptotal}
\end{eqnarray}
with $c$ the concentration of impurities. The factor 2 in the
definition of $T_1$ with respect to the spin-flip time $T_{\rm sf}$
has a historical origin as $T_1$ was derived from the full linewidth
at half-amplitude of conduction elecron spin resonance
spectra.\cite{Monod82,fedorov08} In the literature, $T_1$ is usually
mentioned as the measured quantity, but from the point of view of 
scattering theory it is more natural to use $T_{\rm sf}$.

It is clear that the spin-mixing parameter $b^2$ reflects the host contribution to the spin relaxation while the spin-relaxation time $T_1$ contains also the scattering contribution of the impurity which of course depends on the type of impurity as well as impurity concentration. 
In the present work, we will first discuss the spin-mixing parameter to understand general features of spin relaxation in W(001) thin films. 
After that, by introducing a W adatom on one film surface, we will investigate the spin-relaxation rate $T_1^{-1}$ quantitatively. 
The spin-quantization axis $\sqa$ is at first
chosen perpendicular to the surface of the films, but later we also
examine a variation of the quantization axis revealing anisotropic
effects in spin relaxation. 
The variation of the quantization axis corresponds to an experimental situation where the spin polarization of the injected spin population (or the magnetic field direction in an electron spin resonance experiment) is changed.

We find that the EYP acquires very large values owing to the Rashba effect at the surface and also exhibits an even-odd oscillation with the thickness
of the films following the behavior of the electronic structure of the
surface bands; the same is true for the spin-relaxation rate due to
scattering off W adatoms. 
These effects are the subjects of Sec.~\ref{sec:even-odd} and \ref{sec:Rashba}. 
We then present in Sec.~\ref{sec:anisotropy} the anisotropy of the EYP and of the spin-relaxation rate with respect to the choice of the SQA. 
Finally in Sec.~\ref{sec:remarks} we argue that, as far as the spin relaxation in the surface states is concerned, the mechanism discussed and
calculated here should be dominant over,~e.g., the Dyakonov-Perel
mechanism, prominent in semiconductors or semiconductor
heterostructures where Rashba-type of splitting is also
present\cite{DyakonovPerel,fabian07} but has a much lower magnitude.

Our investigation is based on density-functional calculations within
the local density approximation.\cite{Vosko80} We employ the
full-potential Korringa-Kohn-Rostoker (KKR) Green function
method\cite{Papanikolaou02} with exact treatment of the atomic cell
shapes.\cite{Stefanou90} After a self-consistent full-potential
calculation performed within the scalar-relativistic approximation,
spin-orbit coupling is added when calculating the Fermi surface
properties and the scattering-matrix elements. The formalism for the
calculation is given in detail in Ref.~[\onlinecite{heers11}]. A
similar formalism\cite{footnote} has been applied before in
Refs.~[\onlinecite{gradhand09,gradhand10}] where the results compared
well with experiment.  An angular momentum cutoff of
$l_{\mathrm{max}}=3$ is taken. We use the experimental lattice
constant of bcc tungsten of $a = 3.165$~\AA.  For an interpretation of
our results and a separation of causes we perform in some cases
numerical experiments by switching on and off the spin-orbit coupling
in the system.  The W adatom impurity is treated using the J\"ulich
KKR impurity-embedding code ({\tt KKRimp}) which enables us to treat
with the charge and spin-density self-consistently including the
perturbation of the nearest neighbors of the impurity.  Lattice
relaxations are not taken into account.
 
It should be mentioned that W(001) films have been a subject of
intensive theoretical and experimental reasearch for a long time,~cf.~Refs.~[\onlinecite{weng78,posternak80,mattheiss84}] and citations therein. 
It is known that at low temperatures the W(001) surface undergoes a c($2\times2$) reconstruction, while the ideal structure is restored at higher temperatures.\cite{felter77,debe77}
Additionally, the bcc structure of W(001) films has been extensively
studied owing to its multiple applications as a substrate used for
deposition.\cite{singh86,li89,gaudin94,almanstotter99} 
Recently, because of the strong spin-orbit interaction W(001) is used as subtrate for ultrathin magnetic films generating a strong Dzyaloshinki-Moriya interaction introducing complex magnetic structures.\cite{Ferriani08} 
However, most important for our study are the surface states. 
Together with Cu(111), W(001) is one of the first systems for which surface states were predicted to exist theoretically.\cite{swanson66,waclawski72}

\section{Even-odd effect in spin-mixing parameter and spin-relaxation rate \label{sec:even-odd}}

We first present our results on an unexpected even-odd oscillatory
variation that we observed in the Elliott-Yafet parameter of the film and the 
spin-relaxation rate due to a W adatom with respect to the number of layers in the W(001) film.
\begin{figure}
\includegraphics[scale=0.35,trim= 40 40 10 30,clip=true]{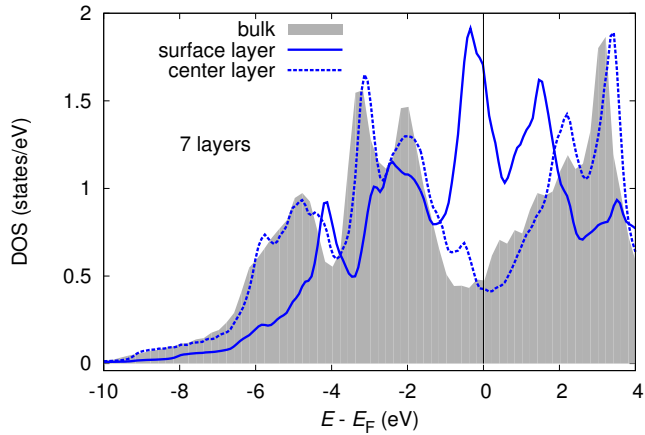} 
\caption{\small (Color online) The atom-resolved density of states (DOS), calculated ignoring the SOC, of the surface- and central layer of a 7-layer thick W(001) film in
  comparison to the DOS of bulk W shown as a grey-shaded area.}
\label{dosW}
\end{figure}   

To start the discussion, we take a look at the comparison of the
density of states (DOS) of W bulk and a typical representative of the
films studied in the following --- a 7-layer W(001) film --- shown
in Fig.~\ref{dosW}. 
In this particular calculation the spin-orbit coupling was not included.  
Characteristic of a transition metal with bcc structure, the bulk DOS shows a bimodular behavior with a dip in the middle separating bonding from antibonding states, which for W lies at the Fermi energy. 
This structure of the DOS is also clearly preserved in the center of the thin W(001) film.
In contrast, the local DOS in the surface atomic layer of the film displays a peak
at $\EF$, which can be attributed to the presence of surface
states. 
Obviously, this should result in a significant surface
contribution to the Fermi-surface-dependent quantities and in
particular the Elliott--Yafet parameter. 
The DOS of W(001) films of different thickness, considered later, are very similar to those in the case of a 7-layer film.

\begin{figure}
\includegraphics[scale=0.37,trim= 1 1 1 1,clip=true,angle=90]{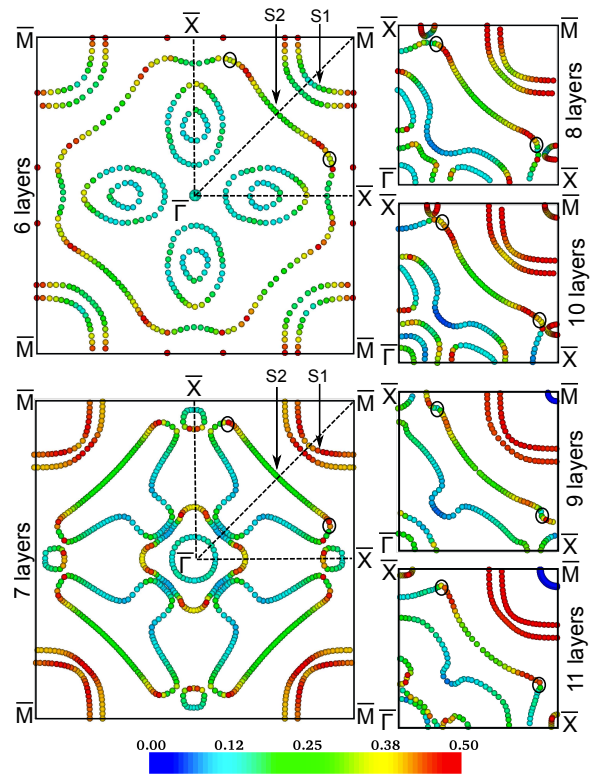} \\
\caption{\small (Color online) Fermi surfaces calculated including SOC and distribution of the spin-mixing parameter $b_{\vc{k}}^2$ (color of the Fermi
  surface points) for W(001) films of various thickness.
The Fermi surfaces are grouped into those for films with even number of layers
  (upper group) and odd number of layers (lower group).  
The surface states S1 and S2 are marked with arrows. 
Small circles indicate where S2 merges into the bulk-like states.}
\label{bsqWcombined}
\end{figure}  

Next, we include the spin-orbit coupling, choose the spin-quantization axis perpendicular to the film ($\hat{s}||[001]$) and calculate the Fermi
surfaces and the distribution of the spin-mixing parameter
$b_{\vc{k}}^2$ for W(001) films of varying thickness.
The results are presented in Fig.~\ref{bsqWcombined}. 
The most important feature for our study are the surface states. These
are indicated as ``S1'' and ``S2,'' and pointed at by arrows in the vicinity of the $\overline{\mathrm M}$ point on the Fermi surface. 
S1 and S2 are present at all film thicknesses and we will discuss them in more detail below. 
Here we merely note for clarity that the S2-line merges into bulk-like states at its two ends (indicated by two small circles in Fig.~\ref{bsqWcombined}), while S1 does not; also, that S1 appears as a double line due to SOC-induced Rashba splitting and due to the interaction between the two film surfaces,
while S2 appears as a single line because its partner is higher than
\EF\ in energy.

\begin{figure}
\includegraphics[scale=0.45,trim=0 250 0 40,clip=true]{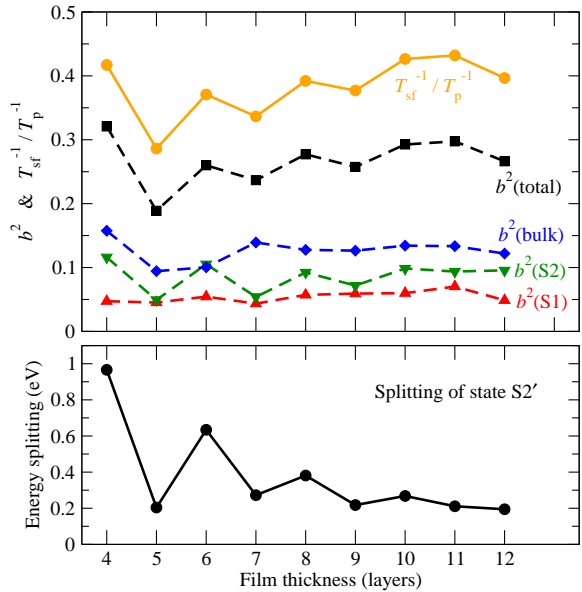}
\caption{(Color online) Top: The dashed lines show the thickness
  dependence of the Elliott--Yafet parameter in W(001) thin films with
  the SQA perpendicular to the film. The total EYP (squares) is
  decomposed into contributions from the surface states S1 and S2
  (triangles) and from the bulk states (diamonds) of the film.  The full line
  (circles) shows the ratio between spin-flip rate and 
  momentum-relaxation rate, $T_{\rm sf}^{-1}/T_{\rm p}^{-1}$ for the case of
  self-adatom scattering [W adatom on W(001)].  The even-odd effect is
  evident in the ratio following very closely the oscillation of the
  Elliott-Yafet parameter.
  Bottom: Splitting of the state S$2'$ (without SOC) due to the interaction between the two surface states of opposite surfaces (see Fig.~\ref{bandW} for the band structure and for the position where the splitting of S$2'$ is calculated).
  The oscillatory behavior of the splitting as function of film thickness
  correlates with the behavior of the EYP and spin-flip rate. The
  lines are guides to the eye.}
\label{fig:bsqW}
\end{figure}

The importance of the surface states for spin relaxation can be
ascertained by looking at the Fermi-surface-integrated Elliott-Yafet
parameter $b^2$ [Eq.~(\ref{bsq})] as a function of film thickness,
additionally decomposed into surface and bulk contributions and 
presented in the top panel of Fig.~\ref{fig:bsqW}.
The latter decomposition was performed by integrating in
Eq.~(\ref{bsq}) over only the surface-state part or only  the bulk-state
part of the Fermi surface, while keeping the denominator $n(\EF)$ fixed
at the value of the total density of states.
It is striking that for all considered film thicknesses the overall contribution of the surface states to the EYP is comparable, or even larger than, the bulk 
contribution. 
It is well known that the amplitude of the surface states drops exponentially as a function of distance from the surface. 
Therefore, in the limit of large film thickness, the relative contribution of the surface states decreases and the bulk contribution becomes prominent. 
However, in the ultrathin films studied here (maximum 12 layers thickness), the surface states are more like quantum-well states and do not fully decay in the center of the film. 
Additionally, there are two more reasons why the bulk limit is not reached fast.
First, the density of bulk-like states at $E_{\rm F}$ is at a minimum as we saw
in Fig.~\ref{dosW}, thus the bulk contribution to $b^2$ sets in only
slowly. 
Second, the Fermi surface of bulk W is rather complex with the
value of $b_{\vc{k}}^2$ strongly varying over it (see
e.g.\ Ref.~\onlinecite{Zimmermann12}), thus many film layers are needed
to achieve the equivalent of a good resolution in the [001] direction
of the bulk Brillouin zone that would yield the bulk limit
$b^2=0.065$.\cite{Zimmermann12} Surprisingly, as we observe from
Fig.~\ref{fig:bsqW}, the contribution of the surface state S2 and correspondingly the
total EYP displays a pronounced even-odd oscillation as a function of film thickness up to 10 layers.

\begin{figure}
\includegraphics[scale=0.40]{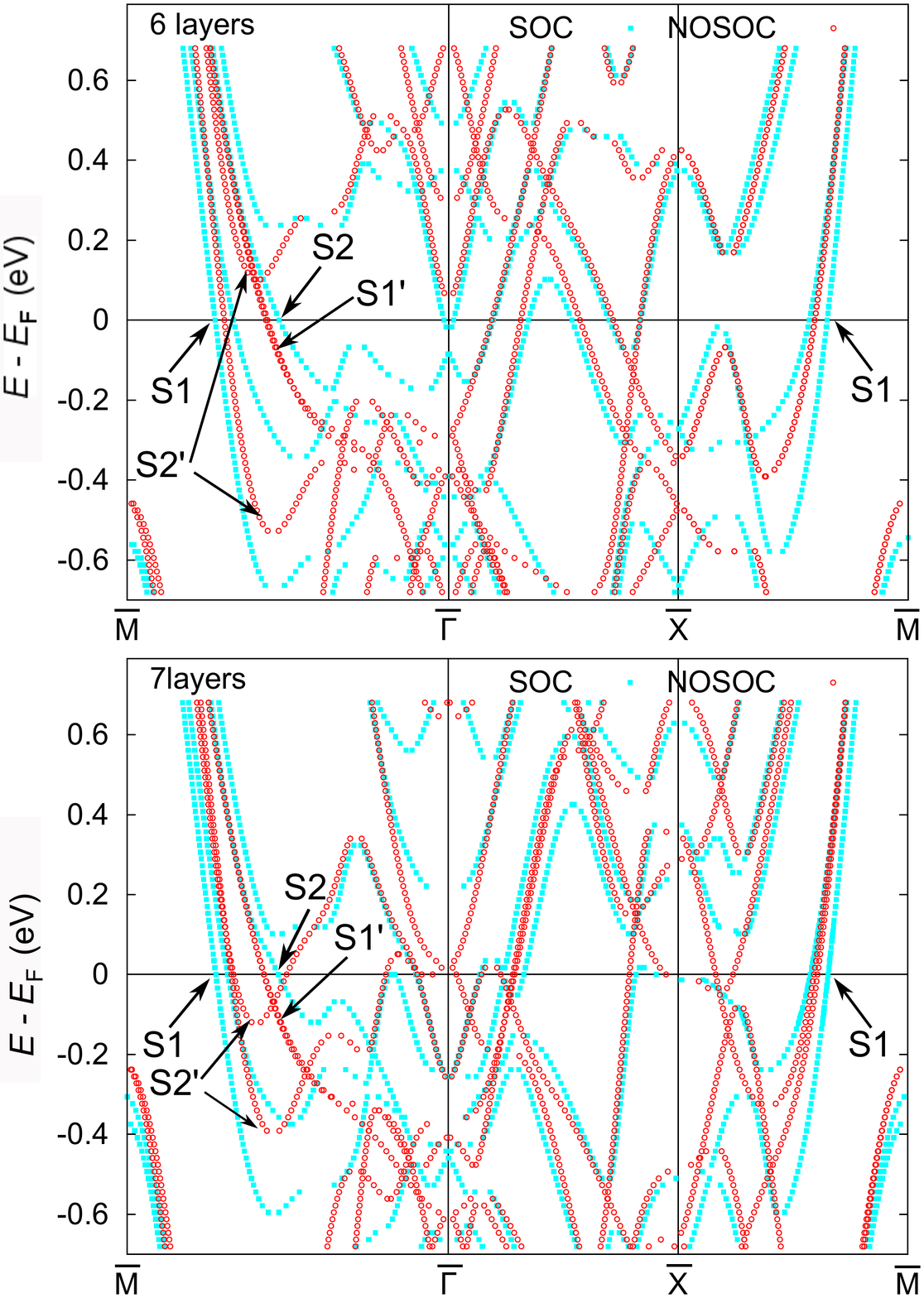} \\
\caption{(Color online) The band structure of 6- and 7-layer W(001)
  films with (filled blue squares) and without (red open circles)
  spin-orbit coupling. Arrows mark the surface states S1 and S2
  corresponding to the calculation with SOC and S$1'$ and S$2'$
  corresponding to the calculation without SOC. An even-odd effect in
  the thickness-dependent splitting of S$2'$ is responsible for the
  even-odd effect in the EYP.}
\label{bandW}
\end{figure}

If we examine the distribution of $b_{\vc{k}}^2$ on the Fermi surface
for several film thicknesses, presented in Fig.~\ref{bsqWcombined}, it
becomes apparent that $b_{\vc{k}}^2$ exhibits large variations in
magnitude, both as a function of the position at the Fermi surface as
well as the film thickness.  
We also clearly observe that among all states at the Fermi surface the largest spin-mixing occurs in the surface states S1 and S2.

Let us try to understand the oscillation with film thickness starting
from the band structure and symmetry properties of the surface states.
We consider the surface states in the absence of spin-orbit coupling because it is simpler to apply symmetry arguments in this case.  We
name the states S$1'$ and S$2'$ where the primed indices are used to
distinguish the case without SOC; S$1'$ and S$2'$ are mixed with each other under the action of the SOC Hamiltonian to produce S1 and S2.

We consider the band structure of S$1'$ and S$2'$ along the
$\overline{\Gamma} \overline{\mathrm M}$ diagonal of the Brillouin zone (red open circles in Fig.~\ref{bandW}).
S$1'$ and S$2'$ comprise mainly $d$ orbitals. 
We set a coordinate system with $x$ and $y$ along the [100] and [010] directions in the surface and $z$ normal to the surface. 
We distinguish two types of $d$ orbitals with respect to their reflection properties about the diagonal [110] ($\overline{\Gamma} \overline{\mathrm M}$):
even [$\frac{1}{\sqrt{2}}(d_{xz} + d_{yz})$, $d_{xy}$, and $d_{z^2}$]
and odd [$\frac{1}{\sqrt{2}}(d_{xz} - d_{yz})$ and
$d_{x^2-y^2}$]. 
Staying on the high-symmetry line $\vc{k}\in\overline{\Gamma} \overline{\mathrm M}$, the Bloch states $\Psi_{\vc{k}}(\vc{r})$ derived from odd orbitals
show nodes
%at the bond center with period $\sqrt{2}a$ for $\vc{r}$ varying along the
%[$\bar{1}$10] direction, 
along the [110] diagonal, which has a consequence of a high in-plane
kinetic energy, not leaving enough energy for penetration into the
bulk region. 
Therefore, the odd-$d$-orbital surface states of the opposite surfaces couple
very weakly to each other and show an almost vanishing splitting
already at small film thicknesses.
On the other hand, the even states, relieved from this nodal structure, have less in-plane kinetic energy and thus enough energy to penetrate into the bulk region and hybridize with their likes of the opposite surface; then the resulting hybrids show a splitting even at larger thicknesses.
This parity-dependence of the surface states was observed in the past by
Mattheiss and Hamann.\cite{mattheiss84} 
The consequence of this can be seen e.g.~in Fig.~\ref{bandW} for the bands without spin-orbit coupling along $\overline{\Gamma} \overline{\mathrm M}$ where the S$2'$ bands show a large splitting whereas the bands S$1'$ show an almost complete degeneracy.

The even-odd effect of the EYP of state S2 can now be traced back to
surface state S$2'$. 
The aforementioned splitting of S$2'$ due to the interaction between the opposite surfaces exhibits an even-odd behavior: for an even number of layers the splitting is large, for odd it is considerably smaller due to an oscillation in the
coupling. 
This can be seen in Fig.~\ref{fig:bsqW} (bottom panel) where the splitting of S$2'$ has been calculated at the position indicated by arrows in Fig.~\ref{bandW}.
One immediately recognizes the striking correlation between the oscillations in the splitting of S$2'$ and in the EYP shown in the same figure. 
We should comment that the origin of the even-odd effect in the coupling could not be fully explained by the symmetry properties of the wavefunctions. 
However, it is clear that owing to the different inversion symmetry center in even- and odd-layer films, the overlap between the opposite-side surface states is different which contributes to the even-odd effect.
%Although the origin of the even-odd effect in the coupling cannot be related to exact symmetry properties of the wavefunctions, still it is clear that the effect originates in the relative shift of the atom centers of the opposite surfaces in the even-layer case, changing the overlap between the opposite-side surface states.

Including the SOC, the even-odd dependence of the energy splitting
will have a profound effect on the spin-mixing parameter, since the
energy distance between states affects crucially the value of
$b_{\vc{k}}$ in Eq.~(\ref{mixwavefunction}). To justify this statement
we remind the reader of Elliott's perturbative
expansion:\cite{elliott54} for the $n$-th band a summation over all
other bands $n'\neq n$ should be performed, reading $b_{n\vc{k}}^2 = \sum_{n'}
|\langle
\Psi^0_{n\vc{k}}|\xi\left(LS\right)^{\uparrow\downarrow}|\Psi^0_{n'\vc{k}}\rangle|^2/
(E^0_{n\vc{k}}-E^0_{n'\vc{k}})^2$ in first order perturbation
theory, where the superscript 0 refers to the wavefunctions and eigenstates without SOC, $\xi$ is the SOC strength and $(LS)^{\uparrow\downarrow}$ is the spin-flip part of SOC operator. 
Since the energy difference of the states appears in the denominator (in all orders of the perturbation expansion), changing the energy splitting will strongly affect the value of $b_{\vc{k}}^2$.
Thus emerges the correlation between the splitting of S$2'$ without SOC and the EYP with SOC in Fig.~\ref{fig:bsqW}.

Now we investigate the spin relaxation due to the W adatom impurities located on one film surface.
In the upper panel of Fig.~\ref{fig:bsqW} we also show the ratio
between the spin-flip rate and momentum-relaxation rate, $T_{\rm
  sf}^{-1}/T_{\rm p}^{-1}$, as a function of the film thickness. 
We see that the magnitude of the ratio is of the same order as the
Elliott-Yafet parameter and that the oscillatory behavior of the two
quantities is clearly correlated, even though Elliott's approximate
relation $T_{\rm sf}^{-1}/T_{\rm p}^{-1}\approx 4b^2$ does not hold. 
The deviation from Elliott's approximation is not surprising since it
holds under the assumptions that $b^2$ is small and that the scattering
is weak enough to be described within first-order perturbation
theory. 
In any case the high values of $T_{\rm sf}^{-1}/T_{\rm p}^{-1}$, of the order of 0.3-0.4, show that approximately every third scattering event includes a spin-flip. 
This high spin-flip rate is certainly related with the fact that a W adatom introduces an additional contribution to spin-flip scattering via its internal
spin-orbit coupling. 
It can be remarked that the ratio between spin-flip rate and momentum-relaxation rate does not depend on the impurity concentration, but the values of the two quantities seperately do. 
In order to have a quantitative estimate, the impurity concentration $c$ in Eq.~\ref{eq:T1k} and ~\ref{eq:tpk} is set to be 1\%. 
This gives us, for example, a value of the spin-relaxation rate of 11.63 ps$^{-1}$/at\% and momentum-relaxation rate of 13.69 ps$^{-1}$/at\% for a 10-layer W(001) thin film by scattering off W adatoms.
We will return to the spin relaxation rate in Sec.~\ref{sec:anisotropy}.

\section{Rashba-character of surface states \label{sec:Rashba}}

\begin{figure}
\includegraphics[scale=0.4,trim=0 350 80 20,clip=true]{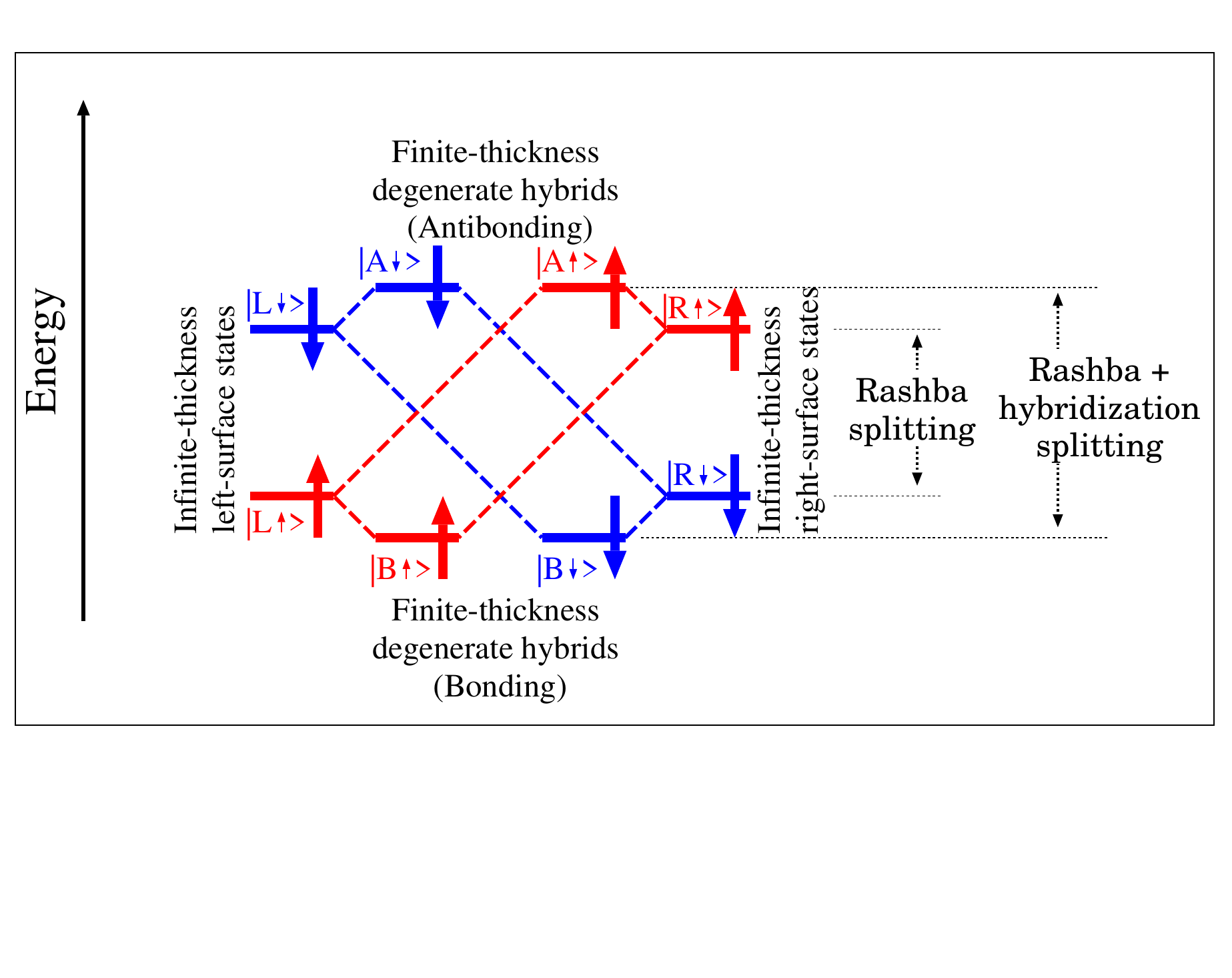}
\includegraphics[scale=0.35,trim=0 0 50 300]{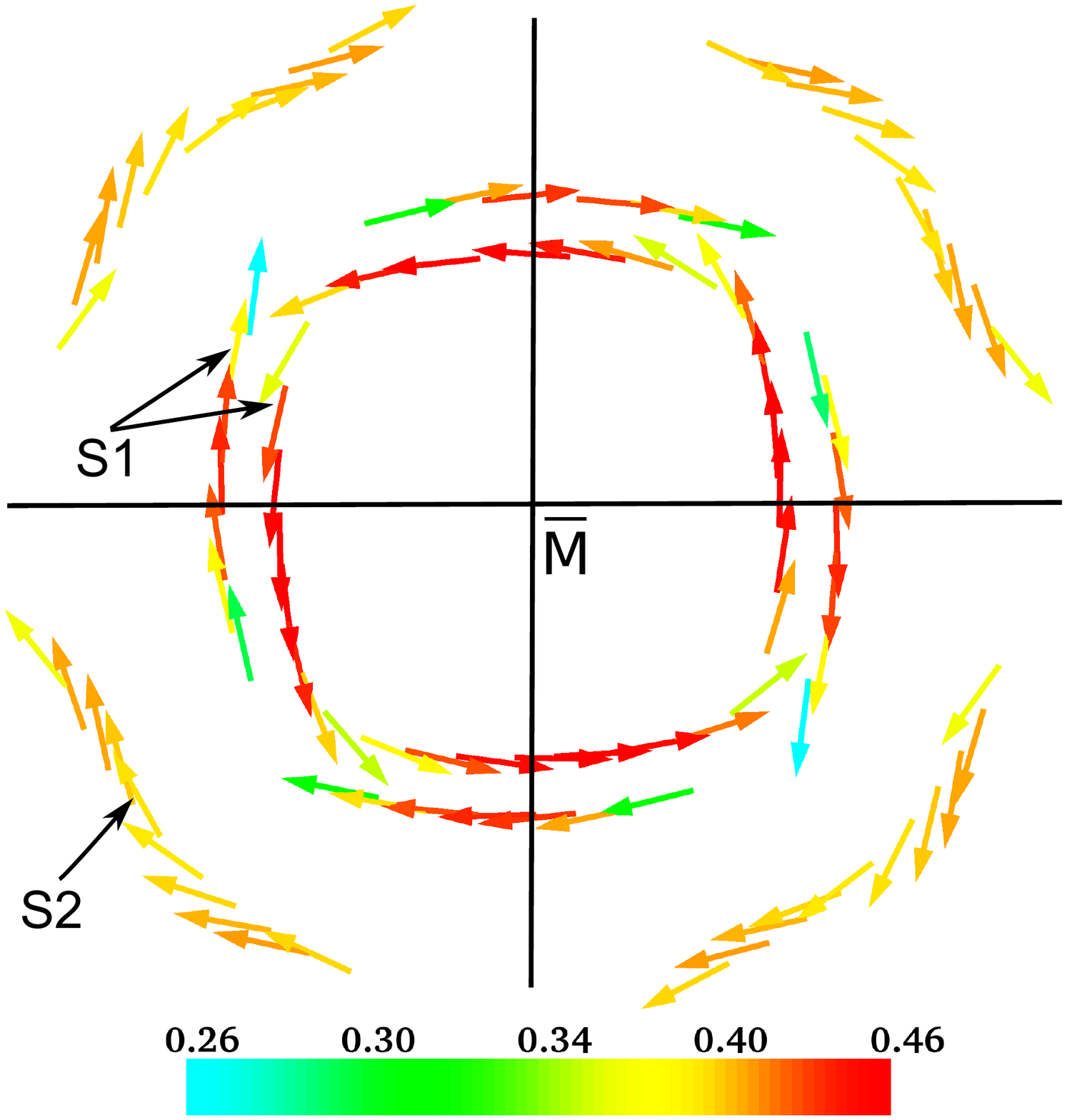}
\caption{(Color online) Top: Schematic diagram showing the energy
  levels of the Rashba states at the film surfaces and their
  hybridization when the film thickness is finite and the states of
  opposite surfaces interact. The full lines represent the energy
  levels before and after hybridization and the dashed lines show
  which level pairs are hybridized. Bottom: The spin polarization of the surface
  states around the $\overline{\rm M}$-point of an 11-layer
  W(001) film. The direction of the spin polarization is given by
  arrows, while its magnitude is given by means of a color code of the
  arrows. Surface states S1 and S2 are marked.
\label{rashba}}
\end{figure}

This section is dedicated to the investigation of the Rashba effect at
the W(001) surface. The essence of the Rashba effect at surfaces lies
in a crystal-momentum dependent splitting of surface states due to the
presence of spin-orbit coupling together with the structural inversion asymmetry caused by the surface.\cite{rashba60,krupin05} 
As was first orbserved experimentally in Au(111) by LaShell {\it et al.}\cite{lashell96} and then theoreticaly explained by Henk and co-workers,\cite{henk03} the surface states of Au(111) display a SOC-derived dispersion relation very similar to that of the two-dimensional electron gas (2DEG) described by a
two-band Rashba model.\cite{rashba60} 
One signature of the Rashba effect is a Fermi surface consisting of two concentric rings with the spin polarization perpendicular to the group velocity but pointing in opposite directions at each of the
rings. 
We will refer to the resulting $\vc{k}$-space spin texture as
\emph{spin-polarization field} in the following (which is closely related to the spin-orbit field defined e.g. in Ref.~\onlinecite{fabian07}).

The Rashba splitting results in a degeneracy lifting where the pairing
implied by Eq.~(\ref{mixwavefunction}) does not hold any more.  Since
the effect originates in the broken inversion symmetry at the surface,
it can be captured computationally by treating half-infinite systems
or by breaking the symmetry between two surfaces of a finite symmetric
slab via deposition on a substrate or adsorption of a monolayer of a different material on one
surface of the slab, e.g.\ a Cu monolayer on an Au film\cite{bihlmayer} or H
adsorption on W(110).\cite{Eiguren09} 
Here, we follow an alternative approach, keeping the inversion
symmetry but projecting out the appropriate state from the doubly
degenerate subspace defined in Eq.~(\ref{mixwavefunction}).\cite{Hirahara08}

We introduce an intuitive view of our approach by means of a schematic energy level
diagram in Fig.~\ref{rashba} (top). We first imagine the film in the limit
of infinite thickness, with the states of each surface not interacting
with the opposite surface. Then, a Rashba splitting occurs due to
SOC. Fixing a $\vc{k}$-point on the surface band, the splitting
results in spin-polarized levels $\bra{\rm L\up}$ and $\bra{\rm L\dn}$
on the left surface (indicated by L) and similarly $\bra{\rm R\up}$
and $\bra{\rm R\dn}$ on the right surface. The Rashba splitting
$\Delta$ is the same in the two surfaces but the order with
respect to energy in which the levels of spin polarization $\up$ and $\dn$
occur is reversed in the two surfaces because of the opposite
direction of the surface normal, according to standard Rashba
theory. Reducing the film thickness, the left and right states
interact via a hopping $t$ (let us assume $t\ll \Delta$) and
hybrid states are formed with bonding (B) and antibonding (A) nature:
$\bra{\rm B\up}\approx \bra{\rm L\up}+\frac{t}{\Delta}\bra{\rm
  R\up}$, $\bra{\rm A\up}\approx -\frac{t}{\Delta}\bra{\rm
  L\up}+\bra{\rm R\up}$. Thus, $\bra{\rm B\up}$ is localized more on
the left surface and $\bra{\rm A\up}$ more on the right
surface. Analogously, hybrids $\bra{\rm B\dn}$ and $\bra{\rm A\dn}$
are formed. A calculation of the film band structure finds linear combinations of the degenerate bonding $\alpha\bra{\rm B\up}+\beta\bra{\rm B\dn}$ as well as antibonding $\alpha\bra{\rm A\up}+\beta\bra{\rm A\dn}$ hybrids; such linear combinations are basically the degenerate
wavefunctions $\Psi_{\vc{k}}^{\pm}$ described by
Eq.~(\ref{mixwavefunction}) in the particular case of surface
states. Returning to the $\Psi_{\vc{k}}^{\pm}$ notation of
Eq.~(\ref{mixwavefunction}), it has then to be decided which
particular linear combination is of interest for the physics of the
problem at hand. For instance, for the calculation of the EYP in the
previous section the linear combinations were chosen so that the spin
expectation value along the $z$ direction was maximized. Here, on the
other hand, we want to choose constants $\alpha_{\vc{k}}$ and
$\beta_{\vc{k}}$ (not depending on $\vc{r}$) in such a way that the
combination $\alpha_{\mathbf k} \Psi^{+}_{\mathbf k} + \beta_{\mathbf
  k} \Psi^{-}_{\mathbf k}$ resembles as much as possible the Rashba
states of the infinite-thickness film; i.e., in a way that the
wavefunctions $\bra{\rm B\up}$, $\bra{\rm B\dn}$, $\bra{\rm A\up}$,
and $\bra{\rm A\dn}$ of Fig.~\ref{rashba} are
retrieved. For this reason we pick one surface, say the left, and we
find the linear combination that maximizes the spin expectation value
within the particular surface atomic layer and call the resulting
wavefunction $\Psi^{\rm max, L}_{\mathbf k}=\alpha_{\mathbf k}
\Psi^{+}_{\mathbf k} + \beta_{\mathbf k} \Psi^{-}_{\mathbf k}$. We
thus define the polarization in the surface layer as
\begin{equation}
 \langle S^i_{\mathbf k}\rangle_{\rm surf}=\frac{1}{2}\int\limits_{\rm
   surf. layer}\left[\Psi^{\rm max, L}_{\mathbf k}({\mathbf r})\right]^{\dagger}
  \sigma^i\,\Psi^{\rm max, L}_{\mathbf k}({\mathbf r})\,d^3r,
\label{ssurface}
\end{equation} 
$i\in \{x, y, z\}$, and demand that $\alpha_{\vc{k}}$ and
$\beta_{\vc{k}}$ are such that $|\langle \vc{S}_{\mathbf
  k}\rangle_{\rm surf}|=\sqrt{\langle S^x_{\mathbf k}\rangle_{\rm
    surf}^2+\langle S^y_{\mathbf k}\rangle_{\rm surf}^2+\langle
  S^z_{\mathbf k}\rangle_{\rm surf}^2}$ is maximized.  Then we observe
that $\Psi^{\rm max, L}_{\mathbf k}$ has both the spin and charge
density more localized on the left surface (numerical result not shown
here explicitly).  Since the film potential in our calculation is
still inversion-symmetric, $\Psi^{\rm max, L}_{\mathbf
  k}=\alpha_{\mathbf k} \Psi^{+}_{\mathbf k} + \beta_{\mathbf k}
\Psi^{-}_{\mathbf k}$ has an orthogonal degenerate partner, $\Psi^{\rm
  max, R}_{\mathbf k} = \alpha^*_{-\mathbf k} \Psi^{-}_{\mathbf k} -
\beta^*_{-\mathbf k} \Psi^{+}_{\mathbf k}$, which is more localized on
the opposite surface and has the opposite spin expectation value. We
say ``more localized'' and not ``completely localized'' because
$\Psi^{\rm max, L/R}_{\mathbf k}$ are still left-right hybrids but
each has higher amplitude on its representative surface. Thus, we have
conveniently separated the degenerate surface states in a way that
they naturally evolve into the single surface state case in the limit
of infinite film thickness. It should be stressed here that maximizing
$|\langle \vc{S}_{\mathbf k}\rangle_{\rm surf}|$ is only the means of
choosing a reasonable approximation to the single-surface
state. However, we analyze the resulting spin-polarization field by
calculating the spin expectation value of $\Psi^{\rm max, L}_{\mathbf
  k}$ over all layers, $\langle \vc{S}_{\mathbf k}\rangle =
\frac{1}{2}\langle \Psi^{\rm max, L}_{\mathbf k}|\boldsymbol{\sigma}|
\Psi^{\rm max, L}_{\mathbf k}\rangle$.

In Fig.~\ref{rashba} we present our results on the spin-polarization field,
represented by arrows, for an 11-layer W(001) film. The figure is
focused on the area of reciprocal space around the $\overline{\rm M}$-point, i.e.,
partly outside the first Brillouin zone, to show the 
contours which correspond to the surface states S1 and
S2. The starting point of each arrow corresponds to a $\vc{k}$-point on the
Fermi surface, the direction of the arrow to the
direction of $\langle \vc{S}_{\mathbf k}\rangle$, and the color
code to its magnitude.
The out-plane component of spin polarization is found to vanish zero.

The spin-polarization field of surface state S1 (two concentric rings
around $\overline{\rm M}$) reminds one of the ``pure'' Rashba states at the Au(111) surface, in the sense that $\langle \vc{S}_{\mathbf k}\rangle$
is in-plane and almost perpendicular to $\vc{k}-\vc{k}_0$,
where $\vc{k}_0$ is the ring center (here the $\overline{\rm M}$-point; in the Rashba
model for Au(111) the $\overline{\Gamma}$ point). More precisely we observe that $\langle\vc{S}_{\mathbf k}\rangle$ is perpendicular to the group velocity,
i.e., $\langle\vc{S}_{\mathbf k}\rangle$ is along the Fermi-surface tangent, as expected from the Rashba model. 
As for the spin-polarization field of the state S2, first we
should remind the reader that S2 merges with the bulk-state
continuum (see Fig.~\ref{bsqWcombined}), excluded from the plot, which
leads to the fact that the shown contour in Fig.~\ref{rashba} ends
seemingly abruptly. Second, S2 also has a partner with opposite
direction of the spin-polarization field, which lies higher in energy
(see the band-structure in Fig.~\ref{bandW}).  
$\langle\vc{S}_{\mathbf k}\rangle$ in S2 is also in-plane and perpendicular to the group velocity.

Finally, it should be noted that the magnitude of the
spin-polarization fields, $|\frac{1}{2}\langle \Psi_{\mathbf
  k}(\vc{r})|\boldsymbol{\sigma}|\Psi_{\mathbf k}(\vc{r}) \rangle|$
cannot reach the maximal value $\frac{1}{2}$, except perhaps in the pure
Rashba model, since an interaction with other states at other energies
at the same $\vc{k}$ via the spin-orbit operator will always be
present and will reduce the value. To view this from a different
perspective, the states $\Psi^{\pm}_{\mathbf k}(\vc{r})$ and their
linear combination $\Psi^{\rm max}_{\mathbf k}(\vc{r})=\alpha_{\mathbf
  k} \Psi^{+}_{\mathbf k}(\vc{r}) + \beta_{\mathbf k}
\Psi^{-}_{\mathbf k}(\vc{r})$ result in a non-collinear spin-density
$\frac{1}{2} \Psi_{\mathbf
  k}(\vc{r})^{\dagger}\boldsymbol{\sigma}\Psi_{\mathbf k}(\vc{r})$
that can be brought in a diagonal form only in an $\vc{r}$-dependent
reference frame; but in order to achieve $|\langle \vc{S}_{\mathbf
  k}\rangle|=\frac{1}{2}$, this reference frame would have to be independent
of $\vc{r}$.

\section{Anisotropy of the Elliott-Yafet parameter and of  the spin-flip
  scattering rate off adatoms\label{sec:anisotropy}}

\begin{figure}[t!]
\begin{center}
\includegraphics[scale=0.35,trim= 0 0 200 100,clip=true]{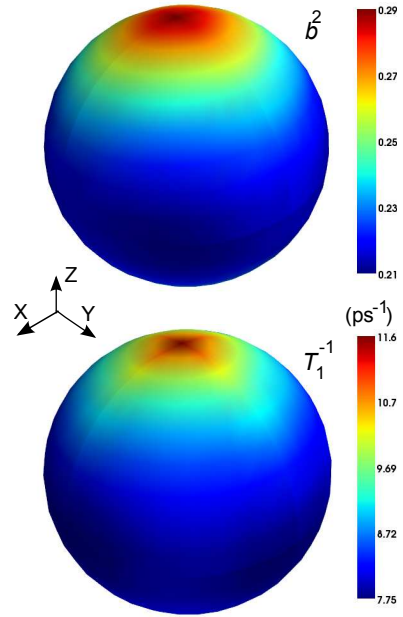}
\caption{(Color online) Value of $b^2(\sqa)$ (top) and
  $T_1^{-1}(\sqa)$ (in units of ps$^{-1}$/at\%) for \sqa\ on 
  the unit sphere for the case of a 10-layer W(001) film with a W
  adatom as scattering defect. The highest spin-relaxation values
  are found for \sqa\ out-of-plane (here taken as the $z$-axis).  
 \label{fig:aniso_int}}
\end{center}
\end{figure}

We now examine the anisotropy of the spin-relaxation rate with respect
to the polarization direction of the injected spin population (i.e.\
the SQA $\sqa$) relative to the crystallographic axes. 
For bulk systems we have already discussed the effect in
Ref.~\onlinecite{Zimmermann12}, pointing out that the reduced symmetry
of thin films will clearly play a role in this phenomenon.
To summarize the origin of the effect,\cite{Zimmermann12} even though the
spin-orbit operator $\vc{L}\cdot\vc{S}$ is independent of the SQA, its
matrix elements are not. 
Particularly relevant here are the matrix elements of the spin-flip SOC $(LS)^{\uparrow\downarrow}$, $\frac{1}{2}(L_+S_- + L_-S_+)$, that are responsible for the
spin-mixing parameter $b_{\vc{k}}^2$ and for the spin-flip transitions
in general. Due to this dependence on the SQA, and considering the
crystallographic symmetry of the W(001) film, we expect at least three
inequivalent directions of \sqa\ for which the spin-relaxation rate
$T_1^{-1}$ will become extremal: perpendicular to the film, in-plane
in the [100] direction and in-plane in the [110] direction. 
Comparing the values of $T_1^{-1}(\sqa)$ in all directions we obtain the
definition of the anisotropy
\begin{equation}
\mathcal{A}[T_1^{-1}]=[\max_{\sqa}T_1^{-1}(\sqa) - \min_{\sqa}T_1^{-1}(\sqa)] / \min_{\sqa}T_1^{-1}(\sqa)
\label{eq:anisotropy}
\end{equation}
which is a somewhat different quantity compared to $\mathcal{A}[b^2]$
introduced in Ref.~\onlinecite{Zimmermann12}, where we had $b^2(\sqa)$
in the place of $T_1^{-1}(\sqa)$ implying the Elliott
approximation. 
We wish to point out here that not only the spin-flip rate but also the spin-conserving rate depends on the SQA, and that part of the anisotropy comes from the SQA dependence of $b_{\vc{k}}^2$ in the Bloch states, while another part comes from the spin-flip scattering off the impurity potential. 
The \emph{total} scattering rate (spin flip plus spin conserving) defined by Eqs.~(\ref{eq:tpk}) and (\ref{eq:tptotal}) is independent of \sqa.

In Fig.~\ref{fig:aniso_int} we show in a color code the value of
$b^2(\sqa)$ and $T_1^{-1}(\sqa)$ as a function of the direction \sqa\
on the unit sphere for a 10-layer W(001) film with a W adatom as
scatterer. 
The maximum spin-relaxation rate of $T_1^{-1}=11.63$ ps$^{-1}$/at\% is obtained for \sqa\ out of plane; the minimum, $T_1^{-1}=7.91$ ps$^{-1}$/at\%, for \sqa\
along the [110] axis. 
This yields an anisotropy of $\mathcal{A}[T_1^{-1}]=47\%$. 
The map of $b^2(\sqa)$ also shows a clear maximum for \sqa\ out of plane with $b^2=0.294$ while in the [110] direction it has a value of $b^2=0.215$ giving an anisotropy of $\mathcal{A}[b^2]=37\%$. 
Evidently there is no complete quantitative correlation between $b^2(\sqa)$ and $T_1^{-1}(\sqa)$, since the Elliott approximation is too crude in this case, but qualitatively the correlation is obvious.  
It should be noted that the in-plane variance of either $b^2(\sqa)$ or $T_1^{-1}(\sqa)$ is small (on the order of 2\%), owing to the high symmetry of the fourfold crystallographic axis.
The anisotropy here for the thin films has a significant higher value compare to the anisotropy of $b^2$ of about 6\% in bulk W, as we have found previously,\cite{Zimmermann12} due to the reduced symmetry and to a great extent due to the surface states

\begin{figure}
\begin{center}
\includegraphics[scale=0.4,trim= 0 0 20 120,clip=true]{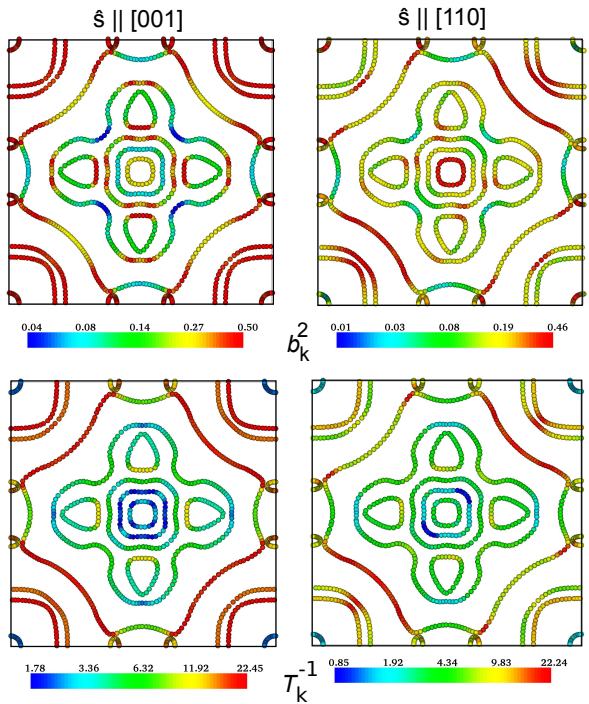}
\caption{(Color online) Distribution of the spin-mixing parameter
  $b_{\vc{k}}^2$ (top) and spin-relaxation rate $T_1^{-1}(\vc{k})$
  (bottom) shown in a color-code (in units of ps$^{-1}$/at\%) on the
  Fermi surface of a 10-layer W(001) film. For the calculation of
  $T_1^{-1}(\vc{k})$, scattering off W adatoms was considered. For the
  calculations shown in the left panels the spin-quantization axis was
  taken in the [001] direction (i.e., perpendicular to the film), for
  the calculations shown in the right panels it was taken in the [110]
  direction (in-plane). The distributions on the Fermi surface are
  clearly compatible with the specific crystal symmetry operations
  that either leave \sqa\ unchanged or result in
  $\sqa\rightarrow-\sqa$. Especially on the surface states the values
  of $b_{\vc{k}}^2$ and $T_1^{-1}(\vc{k})$ are highest in the regions
  where the spin-polarization fields (Fig.~\ref{rashba}) are
  perpendicular to \sqa. \label{fig:aniso_kres}}
\end{center}
\end{figure}

We can analyze the effect further by examining the Fermi-surface-resolved $b_{\vc{k}}^2$ and $T_1^{-1}(\vc{k})$ [defined in Eq.~(\ref{eq:T1k})]. Fig.~\ref{fig:aniso_kres} shows these quantities for $\sqa\parallel[001]$ (left) and $\sqa\parallel[110]$ (right). 
Both quantities [$b_{\vc{k}}^2$ and $T_1^{-1}(\vc{k})$] show
a symmetry compatible to the intersection of the symmetry operations
leaving the Fermi surface invariant and the operations leaving the SQA
invariant plus the inversion. It is interesting to see that even
though the system parameters are outside the prerequisites of the
Elliott approximation ($b^2$ is rather large and the scattering off a
transition metal adatom cannot be considered weak), still a
correlation between the ``highs and lows'' of $b_{\vc{k}}^2$ and
$T_1^{-1}(\vc{k})$ is visible in the color code. 
Also, that the surface state regions show high values for both quantities when the spin-polarization fields (see Sec.~\ref{sec:Rashba} and Fig.~\ref{rashba}) are perpendicular to the SQA and lower values
when they are parallel or antiparallel to the SQA. In this sense, and
since there will be a part of the Rashba states with spin-polarization
field perpendicular to the SQA no matter what the choice of the SQA is
(compare with Fig.~\ref{rashba}), the surface states are clearly ``spin drains'' for the system due to the spin polarization fields that are related to the Rashba effect; the worst
case is $\sqa\parallel[001]$, when \emph{all} spin-polarization fields
are perpendicular to \sqa.

\section{Remarks on the relaxation mechanism in the Rashba states \label{sec:remarks}}

Finally we wish to discuss the physical mechanism of
spin relaxation in the presence of Rashba states. In semiconductors or
semiconductor heterostructures it is naturally assumed that the
spin-polarization fields contribute to spin relaxation via the
Dyakonov-Perel mechanism.\cite{DyakonovPerel,fabian07} There, an
electron occupies a state at $\vc{k}$ that is split in energy by only
a little according to an effective Hamiltonian
$H=-\frac{1}{2}\boldsymbol{\Omega}_{\vc{k}}\cdot\boldsymbol{\sigma}$,
thus in a semiclassical picture the electron spin precesses around the
spin-orbit field $\boldsymbol{\Omega}_{\vc{k}}$ with a Larmor
frequency $|\boldsymbol{\Omega}_{\vc{k}}|$, losing memory of its
original direction before being scattered away (within this effective
model the spin polarization field is parallel or antiparallel to the
spin-orbit field $\boldsymbol{\Omega}_{\vc{k}}$). However, for this to
happen it must be assumed that the electron wavepacket has an energy
spread larger than
$|\boldsymbol{\Omega}_{\vc{k}}|$.\cite{MavropoulosFS,heers11} 
This is possible in semiconductors where $|\boldsymbol{\Omega}_{\vc{k}}|$ is
usually small\cite{fabian07} (of the order of 1meV or less, depending
on temperature, doping concentration, etc.), as usually
$|\boldsymbol{\Omega}_{\vc{k}}|\propto k$ with $\vc{k}$ very close to
the conduction band minimum at $\Gamma$. 
As opposed to this, in metal surfaces the Rashba splitting $\Delta_{\EF}$ of surface states at $\EF$ can be large, of the order of 100~meV, e.g.\ $\Delta_{\EF}\approx 200$meV in W (Fig.~\ref{bandW}), 30meV in Cu(111),\cite{heers11} and 150meV in Au(111).\cite{Nicolay01,bihlmayer} 
It is unlikely that a coherent wavepacket excited by e.g.\ an injected spin current or by microwave radiation in CESR should have such a large energy spread, activating the Dyakonov-Perel mechanism, except perhaps if prepared very precisely by an experiment targeting exactly
this. 
Therefore, we consider the Dyakonov-Perel mechanism not
applicable in the case of the Rashba surface states of most metals,
but we cannot exclude it for special cases e.g.\ lighter metals with
significantly weaker spin-orbit coupling or favorable band structure
(e.g.\ in Ref.~\onlinecite{Nicolay01} we see that Ag(111) shows a splitting of only
$\Delta_{\EF}=2$meV due to the shallow surface state).

\section{Summary and conclusions}

In summary, we have investigated  spin-relaxation physics of W(001)
ultrathin films from first principles. 
We observe that the Elliott--Yafet parameter exhibits an even-odd oscillation with respect to the number of layers of the films, which stems from an even-odd
oscillation in the surface electronic structure of the films at and
near the Fermi energy. 
The oscillation is then inherited by the spin-relaxation rate that depends on the Elliott-Yafet parameter.

We have further identified the Rashba character and spin-polarization
fields of the surface states and discussed how they contribute to the
anisotropy of the spin-relaxation rate with respect to the relative
orientation between the spin-quantization axis and the
crystallographic directions. 
The anisotropy values are much higher compared to those in bulk W. 

We believe that our findings are not only particular to W(001) free
standing films but are more general at least for transition metals in the
bcc structure and even when sandwiched between insulators. 
We base this speculation on three considerations: 
First, we performed calculations (not presented in the present paper) of the Elliott-Yafet parameter for Mo(001) films in the bcc structure and found basically
the same oscillatory effect. 
Second, the existence of surface states is, in general, closely connected to the crystal structure; all bcc transition metals will show a dip in the density of states at the center of the $d$-band and in all cases the breaking of translational symmetry at the (001) surface will produce surface states of the character found
here within the gap of the surface-projected bulk band-structure. 
Third, the mechanism just stated is expected to produce interface states in the case that the film is in contact with an insulator, even if the details of the interface band structure can be more complicated in this case.

As an outlook we believe that it is worthwhile to investigate these
effects in a broader family of ultrathin films. 
Such investigations are in progress and will be reported in a future publication.

\section*{Acknowledgments}

We would like to thank Gustav Bihlmayer for discussions
on the Rashba effect and Martin 
Gradhand for discussions on the spin-flip scattering in films.
This work was financially supported by the MO 1731/3-1 project and SPP
1538 SpinCaT programme of the Deutsche Forschungsgemeinschaft, and the
HGF-YIG NG-513 project of the Helmholtz Gemeinschaft.


\begin{thebibliography}{99}
\bibitem{footnote}
In Refs.~[\onlinecite{gradhand09},\onlinecite{gradhand10}] the fully relativistic KKR method is used, solving the Dirac equation during self-consistency, but in the atomic sphere approximation.
\end{thebibliography}
\end{document}